\documentclass[12pt]{article}

\usepackage{arxiv}

\usepackage[utf8]{inputenc} 
\usepackage[T1]{fontenc}    
\usepackage{hyperref}       
\usepackage{url}            
\usepackage{booktabs}       
\usepackage{amsfonts}       
\usepackage{nicefrac}       
\usepackage{microtype}      
\usepackage{lipsum}
\usepackage{multirow}
\usepackage{threeparttable}
\usepackage{graphicx}

\makeatletter
\renewenvironment{abstract}{
    \if@twocolumn
      \section*{\abstractname}
    \else
      \begin{center}
        {\bfseries \Large\abstractname\vspace{\z@}}
      \end{center}
      \quotation
    \fi}
    {\if@twocolumn\else\endquotation\fi}
\makeatother

\title{\bf Reliable and Efficient Long-Term Social Media Monitoring}

\author{
  Jian Cao\\
  Postdoctoral Scholar In Election Integrity and Data Science\\
  Division of Humanities and Social Sciences\\
  California Institute of Technology\\
  Pasadena, CA 91105 \\
  \texttt{jccit@caltech.edu} \\
   \And
 Nicholas Adams-Cohen \\
 Postdoctoral Scholar \\
  Immigration Policy Lab\\
  Stanford University\\
  Stanford, CA 94305 \\
  \texttt{nadamsco@stanford.edu} \\
    \And
   R. Michael Alvarez\thanks{We thank the John Randolph Haynes and Dora Haynes Foundation for supporting some of this research.  We received support for our use of Google Cloud Platform through Google's COVID-19 research program.   We also thank Anima Anandkumar and Anqi Liu for their work with us on related projects.}\\
 Professor of Political and Computational Social Science \\
  Division of Humanities and Social Sciences\\
  California Institute of Technology\\
  Pasadena, CA 91105 \\
  \texttt{rma@caltech.edu} \\
}

\begin{document}
\maketitle

\begin{abstract}
Social media data is now widely used by many academic researchers. However, long-term social media data collection projects, which most typically involve collecting data from public-use APIs,  often encounter issues when relying on local-area network servers (LANs) to collect high-volume streaming social media data over long periods of time.  In this technical report, we present a cloud-based data collection, pre-processing, and archiving infrastructure, and argue that this system mitigates or resolves the problems most typically encountered when running social media data collection projects on LANs at minimal cloud-computing costs.  We show how this approach works in different cloud computing architectures, and how to adapt the method to collect streaming data from other social media platforms.

\end{abstract}






\section{Introduction}

Social media data is now widely used in many studies in computer and social science \cite{klasnjae_elal_2018}. Many of these studies collect short-term cross-sectional samplings of social media data, while others take advantage of free-access social media APIs and attempt to build longer-term time series that monitor discussions and behaviors online. However, in any project that involves long-term social media data collection efforts, there are many potential issues with collecting a reliable and consistent pipeline of social media data when using on local-area network servers (LANs). In this technical report, we begin by discussing some of the issues that we have encountered in our own experience running a long-term Twitter data collection project (which at this point has been ongoing since 2014).  We then present a cloud-based data collection, pre-processing, and archiving infrastructure which we argue mitigates or resolves many of the problems we have encountered, at minimal cloud-computing costs.   

\section{Problems collecting streaming social media}

If a researcher is interested in quickly collecting cross-sectional social media data from Twitter, the use of the so-called ``streaming'' and ``REST'' APIs are relatively straightforward \cite{russel_2014, steinert_2018}.   Subject to rate limits, Twitter allows researchers to get access to a great deal of incoming Twitter data, including the content of a message, associated metadata, and information about the user account.  In our application, where we are interested in studying the online conversations concerning different political and social topics, collecting data from the Twitter Streaming API by keyword or hashtag data filtering over a brief window of time is relatively straightforward, and is the methodology that many scholars use in their research \cite{oconner_2010, conover_2012, barbera_2014, Beauchamp_2017, adamscohen_2020}.

This situation becomes more complicated if the research project involves longer-term monitoring of conversations and discussions on Twitter.  For example, one of our ongoing projects involves monitoring Twitter mentions of voter issues during elections, requiring us to collect data continuously in the weeks before, during, and after an election.  Ever since beginning this project in 2014, we have worked to refine and improve our methodology for collecting these data \cite{njadamscohen_etal_2017, alvarez_2020}.  Our process focuses on searching for specific keywords that are associated with topics including election fraud, voting by mail, and registering to vote. In another example of long-term social media monitoring, we are developing methods for collecting Twitter conversations using dynamic keyword selection in situations where the discussion might be rapidly-evolving over long periods of time \cite{liu_etal_2019}. 

In attempting to build long-term, multi-year, social media data collection projects on local machines, several prominent problems emerge. In our experience, simple issues can crop up.  Our early work used Python scripts running on local university servers, connected to local-area networks.  We found these scripts often encountered problems accessing the Twitter APIs, had trouble with network access, or competed with other processes running on the servers.  Good programmers can often test scripts collecting longer-term social media data, identify some of these issues, and pause collection to revise code. \cite{njadamscohen_etal_2017}.  

However, even good programmers will have trouble resolving systems failures. An ideal social media monitor should maximize the amount of data gathered while minimizing the influence of any interruptions. Although a robust script that integrates diagnostic tools can help sustain the program, without solving the underlying system failures and limitations the threat of substantial data loss remains.  

First, relying on local hardware introduces difficult, and in some cases, impossible-to-anticipate system failures. Power outages can knock systems offline, and without a secondary local system in place, data in a time series will be permanently lost. Network instability can also undermine data collection efforts, especially during peak-use hours or if network infrastructure is temporarily down for maintenance. Furthermore, if there is permanent system damage to a local system, it can be difficult, if not impossible, to recover data.  

Second, the kinds of local systems most researchers have access to are not designed for the specific needs of collecting real-time streaming data. Collecting these data requires a system that can quickly and effectively obtain, buffer, process, and store large continuous streams of incoming information. Collecting this type of high-frequency, continuously streaming data involves specific computational considerations, with specialized algorithms designed to best capture these data \cite{babcock_2002}. Without a good system designed for these tasks, processing and saving files can temporarily interrupt the Twitter stream and limit the amount of data gathered. Slow processing that doesn't meet the publishing speed will also be constantly disconnected from the social media streams. Given these interruptions are most likely to occur during periods of heavy Twitter traffic, the censored Twitter data may not be missing at random, with systemic omitted Twitter data from the streaming API potentially biasing the results of a study \cite{morstatter_2014}.

Finally, setting up a LAN can potentially limit future collection efforts. As a collection project naturally expands, computational power, active memory, and storage considerations may change. However, local systems can be difficult, expensive, and time-intensive to upgrade, especially if one needs to address these concerns repeatedly in a multi-year research effort.  On the other hand, for seasonal projects such as election monitoring, local systems can be less efficient as they are difficult to downgrade or temporarily shutdown to cut expenses.



\section{Cloud-based social media monitoring}

In this paper, we present specific solutions to data collection on three popular cloud computing services: Google Cloud Platform (GCP), Amazon Web Services (AWS), and Oracle Cloud. While we illustrate our solution on these three platforms, the methods and processes we outline can be generally applied to other cloud services. While we do not claim here that we are the first to develop this type of workflow, we want to provide details about how our long-term social media data collection solution operates for other researchers to evaluate and utilize in their own work.  Essentially we built a process that solves or mitigates many of the problems using LANs by moving the data collection and pre-processing steps onto cloud-computing platforms.\footnote{To reduce cloud-computing data storage costs, and to make the social media data we collect more readily accessible to our research group, we outline a process of piping the pre-processed data to cheap, secure, and easy-to-use data storage applications (here Google Drive). Note that, budget allowing, all data can be stored on a single platform.}

In the next sections of this paper, we first describe the general workflow of our social media monitoring system before providing detailed guidelines on how the process works in GCP, AWS, and Oracle Cloud.

\subsection{Workflow}

We developed an architecture using cloud resources to tackle the problems discussed in the previous section. This system collects as much social media data as rate limits allow in a stable and failure-tolerant manner.

\begin{figure}
    \centering
    \includegraphics[width = 0.9\textwidth]{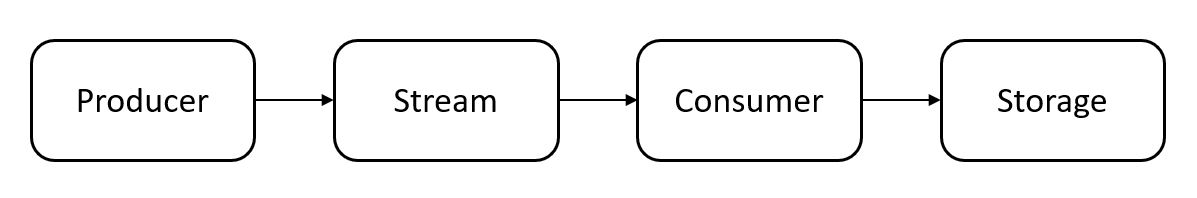}
    \caption{Social Media Monitor Workflow}
    \label{fig:workflow}
\end{figure}

The system consists of four parts: a data producer, a data stream, a data consumer, and storage. As shown in Figure \ref{fig:workflow}, it starts with a producer that requests social media data from the API and acts as a data provider to the other parts of the system. Then the produced data are published in the data stream for temporary storage. The data stream secures every published record on a timeline according to the timestamp. Before the records expire, a data consumer retrieves them from the data stream and either sends them to analytics modules or to the short-term/long-term storage solution.

We describe the workflow of Twitter monitoring step-by-step in the following subsections and will expand the discussion to other social media platforms in the next section.

\subsection{Data Producer}

The data producer runs Python scripts on the cloud compute instances and connects two parts of our data pipeline: the Twitter Stream API and the cloud data stream.

The data producer:
\begin{itemize}
    \item Uses the \texttt{TwitterAPI} package to access the Twitter Stream API, \texttt{google-cloud} to interact with GCP, \texttt{boto3} to interact with AWS, and  \texttt{oci} to interact with Oracle Cloud.
    \item Connects to the Twitter Stream API using Twitter Developer credentials.
    \item Connects to the cloud data stream using credentials/tokens.
    \item Requests tweets from the Twitter Stream API.
    \item Publishes streaming tweets one-by-one to the cloud data stream.
\end{itemize}

We use the Twitter Stream API to collect real-time tweets. The most commonly used Twitter APIs for collecting tweets are the Stream and REST APIs. The Twitter Stream API delivers real-time tweets continuously as soon as they are published and prior to subsequent alterations, such as deletion or censoring by the platform. Furthermore, as tweets are collected right after they are published, they do not carry information regarding the retweets and quotes that occur after a message is sent. On the contrary, the Twitter REST API allows users to search for the tweets that were posted in the past 7 days (30-day and full-archive endpoints are available upon upgrade). These tweets
contain the latest retweet and quote information up to the time of extraction. However, instead of delivering all tweets that meet the filtering rules, the REST API only returns a subset of them that are most relevant. Therefore, to collect maximal tweets, our system focuses on the Stream API, only relying on the REST API as a backup in the case of interruptions.

It is important to note that the rate limits of the Twitter Stream API can cause potential data loss. The Twitter Stream API has a rate limit that only allows the delivery of up to 50 tweets/second. Some projects such as COVID-19 have exclusive unlimited endpoints. Those unlimited endpoints should be prioritized as they have no rate limits.


Selecting a suitable cloud compute instance is also important, as it ensures sufficient computing power at the lowest price. In our COVID-19 project, we use two data producers to request tweets of independent topics. We found that a Python script that produces at Twitter's rate limit (50 tweets/second) occupies on average 20\% of a 2GHz CPU core, and running data producers on separate CPU cores can prevent the processes from competing with each other for resources. For a two-producer project like ours, we suggest using an instance with two CPU cores. We further recommend allocating more than 1GB of active memory space to compile large packages and run data consumer scripts. Table \ref{tab:comparison_specs} shows three popular compute instances in GCP, AWS, and Oracle Cloud that each has specifications that meet the base requirements of the two-producer project. While the AWS system costs less than the others, AWS employs a baseline performance mechanism\footnote{https://docs.aws.amazon.com/AWSEC2/latest/UserGuide/burstable-credits-baseline-concepts.html} that limits the instance's performance once the average CPU usage exceeds the baseline. For the t3.medium instance, performance will be significantly reduced if the user constantly uses more than 20\% of the CPU(s). In general, the performance/price ratios are similar across platforms. Users can always customize or resize their instances to balance the performance and cost.

\begin{table}[h]
    \centering
    \caption{Comparison of Instance Specifications}
    \begin{threeparttable}[t]
        \begin{tabular}{lccc}
        \toprule
        \multicolumn{1}{c}{\multirow{2}{*}{\textbf{Specs}}} & \textbf{GCP Compute}              & \textbf{AWS EC2}       & \textbf{Oracle Compute}   \\
        \multicolumn{1}{c}{}                                & \textbf{VM n2-custom}             & \textbf{EC2 t3.medium} & \textbf{VM.Standard.E2.2} \\ \hline
        \textbf{CPU Series}                                 & \multirow{2}{*}{Intel(R) Xeon(R)} & Intel(R) Xeon(R)       & AMD EPYC                  \\
        \textbf{CPU Model}                                  &                                   & Platinum 8175M         & 7551                      \\
        \textbf{Core Clock}                                 & 2.80GHz                           & 2.50GHz                & 2.00GHz                   \\
        \textbf{Core Count}                                 & 2 Cores                           & 2 Cores                & 2 Cores                   \\
        \textbf{Memory}                                     & 4 GB                              & 4 GB                   & 15 GB                     \\
        \textbf{Baseline Performance}                       & \multirow{2}{*}{100\%}            & \multirow{2}{*}{20\%}  & \multirow{2}{*}{100\%}    \\
        \textbf{(per CPU)*}                                  &                                   &                        &                           \\ \hline
        \textbf{Pricing}                                      & \$50/Month                        & \$30/Month             & \$61/Month                \\ \bottomrule
        \end{tabular}
        \begin{tablenotes}
            \item[*] CPU performance is restricted if average CPU usage exceeds the baseline.
        \end{tablenotes}
    \end{threeparttable}
    \label{tab:comparison_specs}
\end{table}

\subsection{Data Stream}

Cloud data streams are temporary storage services designed specifically for streaming data. Their role in the data collection pipeline is like librarians  -- they receive, organize, and preserve new information from the source and provide multiple means for the users to access the collected information. Data streams are the most essential part of the social media monitor, as they handle heavy data traffic that is beyond most local systems' capabilities, and make each streaming record available to all parts of the system. Data streams improve the reliability of the whole process by making it robust to failures caused by peaks of incoming data and serving as buffers for the subsequent parts of the workflow to re-visit data records if an error should occur.

While the data stream services have different names and specifications in GCP, AWS, and Oracle Cloud, they serve for the same purpose and work in the same way. In the following discussion, we talk about the rate limits, retention periods, and pricing methods of these stream services.

\begin{table}[h]
    \centering
    \caption{Comparison of Cloud Data Streams*}
    \begin{threeparttable}[t]
        \begin{tabular}{ccccc}
        \toprule
        \multicolumn{2}{c}{\textbf{Specs}}                                                                                                                   & \textbf{GCP Pub/Sub}                                                                                             & \textbf{AWS Kinesis}                                                                  & \textbf{Oracle Stream}                                            \\ \hline
        \multicolumn{1}{l|}{\multirow{2}{*}{\textbf{\begin{tabular}[c]{@{}c@{}}Unit Limits\end{tabular}}}} & \multicolumn{1}{c|}{\textbf{Write}} & \multicolumn{1}{c|}{\multirow{2}{*}{None}}                                                                       & \multicolumn{1}{c|}{\begin{tabular}[c]{@{}c@{}}1 MB/s\\ 1,000 Records/s\end{tabular}} & \begin{tabular}[c]{@{}c@{}}1 MB/s\\ Unlimited Writes\end{tabular} \\ \cline{2-2} \cline{4-5} 
        \multicolumn{1}{l|}{}                                                                                          & \multicolumn{1}{c|}{\textbf{Read}}  & \multicolumn{1}{c|}{}                                                                                            & \multicolumn{1}{c|}{\begin{tabular}[c]{@{}c@{}}2 MB/s\\ 5 Reads/s\end{tabular}}       & \begin{tabular}[c]{@{}c@{}}2 MB/s\\ 5 Reads/s\end{tabular}        \\ \hline
        \multicolumn{1}{l|}{\multirow{2}{*}{\textbf{Project Limits}}}                                                  & \multicolumn{1}{c|}{\textbf{Write}} & \multicolumn{1}{c|}{\begin{tabular}[c]{@{}c@{}}50 MB/s (small**)\\ 200 MB/s (large)\end{tabular}}  & \multicolumn{1}{c|}{Unit Limit*N}                                                     & Unit Limit*N                                                      \\ \cline{2-5} 
        \multicolumn{1}{l|}{}                                                                                          & \multicolumn{1}{c|}{\textbf{Read}}  & \multicolumn{1}{c|}{\begin{tabular}[c]{@{}c@{}}100 MB/s (small)\\ 400 MB/s (large)\end{tabular}} & \multicolumn{1}{c|}{Unit Limit*N}                                                     & Unit Limit*N                                                      \\ \hline
        \multicolumn{2}{l}{\textbf{Retention Period}}                                                                                                        & 7 Days                                                                                                           & 24 Hours                                                                              & 24 Hours                                                            \\ \hline
        \multicolumn{2}{l}{\textbf{Pricing}}                                                                                                                 & \begin{tabular}[c]{@{}c@{}}Message Ingestion,\\ Delivery, and Storage\end{tabular}                               & Shards*Hour                                                                           & Partitions*Hour                                                   \\ \bottomrule
        \end{tabular}
        \begin{tablenotes}
            \item[*] Default rate limits can be increased upon upgrade.
            \item[**] Large regions: europe-west1, us-central1, us-east1. Small regions: other regions.
        \end{tablenotes}
    \end{threeparttable}
    \label{tab:comparison_streams}
\end{table}

Table \ref{tab:comparison_streams} shows the default rate limits for the stream services Pub/Sub (GCP), Kinesis (AWS), and Oracle Stream. In the three services, Kinesis and Oracle Stream operate on basic units (called shards in Kinesis, and partitions in Oracle Stream), while Pub/Sub runs as a whole stream.

The basic units are pre-set components of a stream service. They are restricted by default rate limits and are independent of each other. Inside these units, published records are ordered by the timestamps of the ``put" events. A shard of the Kinesis stream supports up to 1,000 records/s or 1 MB/s (whichever is met first) for writes, and 5 requests/s or 2 MB/s for reads. For Oracle Cloud, a partition has similar rate limits to a Kinesis shard, except it has no restrictions on the number of write requests per second.

Users can create more units in their stream services to meet increased demand. In AWS Kinesis, a Python script can use metrics from AWS Cloudwatch to monitor the usage of Kinesis shards, and scale up the number of shards if thresholds are met. For example, to prevent the incoming tweets from hitting the writing rate limits and causing data loss, we can use upper bounds 800 KB/s and 800 records/s as signals for immediate shard-creation. Once the average incoming tweets exceed one of the thresholds, the Python script immediately creates a new shard and lowers the average burden. We can also set lower bounds 500 KB/s and 500 records/s as signals for delayed shard-deletion. Once the average records per shard fall below the lower bounds, we want to delete one or more shards to reduce cost. To be conservative in deleting shards, we recommend waiting at least three hours after the lower bounds signals were triggered. The records that were published in the deleted shards will be available until expiration. For Oracle Cloud, it is not possible to add new partitions to existing streams, so users need to plan ahead or manually migrate to a new stream that has more partitions if incoming record surges.

The GCP Pub/Sub does not involve stream units explicitly. Instead of focusing on unit limits, Pub/Sub users must pay attention to project and resource limits. For a project that is located in large regions (europe-west1, us-central1, us-east1), all Pub/Sub topics combined cannot exceed 200 MB/s for writes and 400 MB/s for reads. The corresponding rate limits in other regions are 50 MB/s and 100 MB/s. The users should create separate GCP projects for large social media monitors that can potentially exceed the project limits.

The retention period determines how long a record is available to be read after being published in the stream. The default retention periods for Kinesis and Oracle Stream are both 24 hours, and it is 7 days for Pub/Sub. Users can upgrade to a longer retention period when creating the stream service.

There are two pricing methods, fixed pricing per unit*hour and flexible pricing per volume of data transmission and storage. Kinesis and Oracle Stream use the former and Pub/Sub uses the latter. The fixed pricing method leaves it to users to optimize the size and cost of the stream, and there are inevitable resource losses in running the units below the rate limits and in re-sizing the stream. On the contrary, flexible pricing only charges the resources being used. However, we wish to emphasize that flexible pricing does not definitively lead to a lower overall cost.\footnote{Please refer to the pricing pages and the cost calculators for each cloud service to better estimate prices.}

\subsection{Data Consumer}

Data consumers are cloud services or customized programs that read records from the data streams and send them to either cloud/local databases for storage or to analytical modules for data processing and analyses. In our social media monitors, we use DataFlow in GCP, Firehose in AWS, and the ``get\_messages" function in Oracle Cloud to extract records from the data stream. We prefer using a cloud service like DataFlow and Firehose instead of relying on customized programs given these services tend to have lower latency and higher stability. To add one more layer of safety, we send the extracted data immediately to cloud databases for short term storage and subsequently invoke BigQuery (GCP), Lambda Functions (AWS), or Data Analytics (Oracle Cloud) for analyses.

\subsection{Storage}

Our social media monitoring system puts collected data temporarily in cloud storage before archiving data in Google Drive folders. Cloud storage services, such as Cloud Storage (GCP), S3 (AWS), Object Storage (Oracle Cloud) are natural data transfer and storage solutions between stream services and compute instances. They are ideal for frequently used data but are not cost-efficient for large sets of raw social media data collected over a long period of time. After the data collection pipeline is checked and the real-time analyses are completed, we store the data summaries and results of the analyses in a cloud MariaDB database and transfer the raw data to a Google Drive shared with all members of the research team. This three-layer storage structure (cloud data stream--cloud storage--Google Drive) is robust to system failures given it inherits the stability and compatibility from the cloud services, and, if an error should happen, the three-layered system can easily recover data from a previous step.

\section{Other Social Media Platforms}

Of course, Twitter is not the only source of dynamic social media data for researchers. While the Twitter API's relative ease of use and open policy make it one of the most popular sources of data in academic studies, many researchers conduct research utilizing data from other popular social media platforms with APIs, such as Reddit\cite{tan2016winning,Nithyanand2017}, YouTube \cite{burgess_2016,rieder_2018}, and Facebook \cite{rieder_2015,kalsnes_2016}.

Our approach can be adapted to collect streaming or dynamic social media data from these other platforms. Considering Figure \ref{fig:workflow}, the main modification in our process would be in the \textbf{Producer} step of the workflow. In this step, we describe setting up a virtual compute instance capable of running code designed to interact with the Twitter API. To modify our process to collect other forms of streaming data, we would simply rewrite this code to interact with another platform's API.

For example, suppose we wished to collect data over time from a particular Reddit community (a subreddit). We could reuse the same process described in the previous section, revising the set of scripts in the Producer step to access data from the Reddit API. There might also be some minor changes required in the code used in the preprocessing and consumer steps, as the data stream from the Reddit API may differ from the Twitter API.  These minor alterations aside, conceptually our approach is highly adaptable for collecting streaming and dynamic data from a variety of social media platforms, as long as they have a public API or their data can easily be obtained via a script that can run on a cloud computing instance.  

\section{Discussion and Conclusion}

Many research groups are using social media data in their studies of political, social, and economic attitudes and behavior.  Interest is increasing in the development of longer-term datasets that can be used to analyze changes over time in attitudes and behavior \cite{adamscohen_2020}.  However, in our efforts to collect longer-term social media datasets from local servers using local-area networks, we have encountered important limitations in the availability and reliability of those systems for these purposes.

To improve the reliability of our longer-term social media data collections process, we have developed a cloud-based infrastructure that is adaptable to several platforms. By moving the data collection and pre-processing stages into the cloud, we  avoid many of the problems encountered when relying on local servers and LANs.  We designed our process to make data easily accessible to our research group by moving the data to a secure and usable storage solution like Google Drive.  We have shown that these processes work across different cloud computing systems, and can be used to collect both Twitter and other streaming social media data.  

There are many opportunities for researchers who are interested in using social media data to study the longer-term dynamics of political, social, and economic attitudes and behavior.  We hope that by providing the details of our framework, other researchers can evaluate and potentially use this report as initial guidance in adopting a cloud-based process, improving their ability to collect similar datasets.


\bibliographystyle{unsrt}  


\end{document}